# Inverse design of higher-order photonic topological insulators

Yafeng Chen[1,2], Fei Meng[2], Yuri Kivshar[3], Baohua Jia[2], Xiaodong Huang[2,1,*]

[1]Key Laboratory of Advanced Technology for Vehicle Body Design & Manufacture, Hunan University, Changsha, 410082, China

[2]Centre of Translational Atomaterials (CTAM), Faculty of Science, Engineering and Technology, Swinburne University of Technology, Hawthorn, VIC 3122, Australia

[3]Nonlinear Physics Center, Research School of Physics, Australian National University, Canberra, ACT 2601 Australia

**Abstract**

Topological photonics revolutionizes some of the traditional approaches to light propagation and manipulation, and it provides unprecedented means for developing novel photonic devices. Recently discovered higher-order topological phases go beyond the conventional bulk-edge correspondence for photonic crystals and introduce novel opportunities for topological protection. Here, we introduce an intelligent numerical approach for inverse design of higher-order photonic topological insulators with great flexibility for controlling both topological edge and topological corner states. In particular, we consider the second-order photonic topological insulator and design several structures supporting both edge and corner states at different frequencies. By carefully programming these structures, we suggest a novel approach for topological routing of the edge and corner states by changing the operational frequency. Our finding paves the way to integrated topological photonic devices with novel functionalities.

* Corresponding author: X. Huang, Tel: +61-3-92175633 , Fax: +61-3-92148264
E-mail: xhuang@swin.edu.au

**1. Introduction**

The discovery of topological phases opened up many new horizons for condensed matter physics [1-3]. Topological insulators with robust gapless edge states can appear in a broad range of applications from spintronics to quantum computation [4-6]. The concept of topological phases has also entered the realm of photonics [7], which revolutionized some of the traditional views on light manipulation and created novel photonic devices, such as backscattering-immune sharply bent waveguides [8], spin-polarized switches [9], robust delay lines [10] and non-reciprocal devices [11].

While most topological systems studied so far have been characterized by the presence of topological states with dimensionality one lower than that of the system, recently, a new class of topological systems, the so-called higher-order topological insulators, has been proposed in electronics [12-14], and further extended to photonic systems [15, 16]. In contrast to conventional topological insulators, higher-order topological insulators support topological states two or more dimensions lower than the system, known as higher-order topological states. One example is the distorted graphene-like photonic lattice that was shown to host lower-dimensional localized states. Another example is that of quadrupole topological insulators, which have been recently implemented in photonic systems and electrical circuits.

Importantly, high-order topological insulators with corner and hinge states deepened our understanding of the bulk-boundary correspondence. In particular, the second-order corner states in 2D photonic crystals (PCs) [17-19] have been experimentally observed [19]. The photonic-crystal nanocavity based on a topological corner state has been demonstrated to support high $Q$ resonances [20].

The topological edge and corner states originated from the designated distribution of dielectric materials within the unit cells of non-trivial and trivial PCs. So far, the design of second-order photonic topological insulators (SPTIs) mainly relies on the trial-and-error process. Even successfully, the resulting property of the topological insulators may be far away from optimum. For instance, the overlapped band gap of nontrivial and trivial PCs is too narrow resulting in a narrow frequency range operated in the design of optical components. Besides, the narrow band gap hardly produces strongly localized edge and corner states, which are obligatory for many practical applications. Therefore, there is an urgent need to formulate the design problem, which can be treated with topology optimization [21-24]. Meanwhile, topology optimization also enables to adjust the operate frequencies of edge and corners states beyond the traditional way through adjusting the size of unit cells.

In this paper, we present an intelligent approach for creating SPTIs using topology optimization. The created SPTIs optimally possess a nearly flat edge state and highly localized corner states. By tuning the operate frequencies of edge and corner states in topology optimization, the optimized SPTIs enable to manipulate lights with different frequencies. Programming these newly created SPTIs, we first realize routing topological edge and corner states with different frequencies.

**2. Inverse-design approach**

Topology optimization employs a powerful numerical algorithm to seek the layout of materials within design domain, so as to minimize or maximize the defined objective function [25]. This concept can be extended to the design of photonic crystals [26-31]. Here, we will extend topology optimization for creating nontrivial and trivial PCs of SPTIs. Consider a 2D PC with $C_{6v}$ symmetry under transverse magnetic (TM) mode, its unit cell is discretised with $N$ finite elements. The dielectric material is selected as silicon with the relative permittivity, $\varepsilon = 12$. The lattice constant is selected as $a = 10$ mm.

Topology optimization firstly constructs a type-I PC, which has two dipolar modes locating above two quadrupolar modes in the band diagram. The detail formulations of topology optimization can refer to the Supplemental Material Sec. 1. By gradually modifying the target frequencies of dipolar modes and quadrupolar modes, we realize a PC (named as NP1) with a wide bandgap between 13.00 GHz and 18.59 GHz, as shown in Figs. 1(a) and (b). The inserts shown in Fig. 1(b) illustrate that the two dipolar modes, $p_x/p_y$, locate above the two quadrupolar modes, $d_{x^2-y^2}/d_{xy}$, at $\Gamma$. Such a parity-inversed band order indicates the bandgap is nontrivial [9, 11]. Next, topology optimization creates a type-II PC (named as TP) by maximizing the bandgap between the second band and the third band [32]. The optimized unit cell and its band diagram are shown in Figs. 1(c) and (d), respectively. It is noticed that the bandgap of the TP ranges from 13.03 GHz to 17.50 GHz.

In order to characterize the topological properties of the optimized PCs, the topological invariant based on the 2D Zak phase can be calculated by the integration of the Berry connection over the first Brillouin zone as,

$$\mathbf{P} = \iint_{\text{1st BZ}} dk_1 dk_2 \mathbf{Tr}(\mathbf{A}), \qquad \mathbf{A} = i\langle u_m | \partial_\mathbf{k} | u_n \rangle, \qquad (1)$$

where $m$ and $n$ run over all bands below the gap. $|u_m\rangle$ is the Bloch function for the $m$th band. Because of the $C_{6v}$ symmetry, the Zak phase can be simplified with

$$P_i = \pi \left( \sum_n q_i^n \mod 2 \right), \qquad (-1)^{q_i^n} = \frac{\eta_n(\mathrm{M})}{\eta_n(\Gamma)} \qquad (2)$$

where $i = x, y$ stands for the direction, and $\eta_n$ denotes the parity at the high symmetry point for the $n$th band. The parity for the calculation of the 2D Zak phase is labelled in the band diagrams. Derived from Eqn. (2), the 2D Zak phase for NP1 equals $(\pi, \pi)$, which ascertains that NP1 is in a topological nontrivial phase. The 2D Zak phase for TP is $(0, 0)$, which confirms that TP is in a topological trivial phase. It is noted that the overlapped bandgap between the nontrivial and trivial PCs is 29.3%, more than two times of the intuitive SPTIs in refs. [17, 18, 20]. Such a wide bulk bandgap enables to cause highly localized edge states and corner states, and provides the programming room for potential applications.

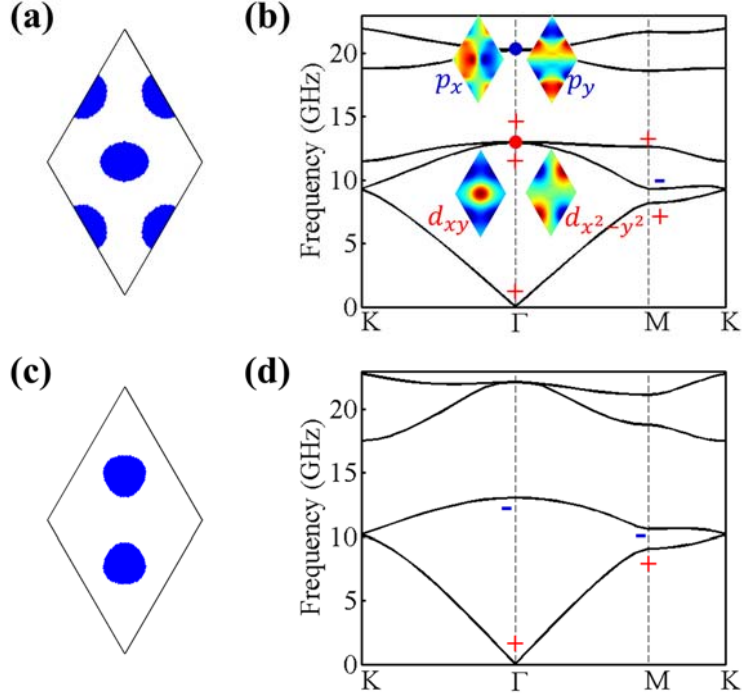

FIG. 1. The optimized primitive unit cell for topological (a) type-I PC and (c) type-II PC. (b, d) Band diagrams for topological nontrivial PC and trivial PC, respectively. The inserts are electric fields profiles of dipolar and quadrupolar modes at the $\Gamma$ point.

**3. Topological edge and corner states**

According to the bulk-boundary correspondence principle [33], topological edge states emerge at the interface between the trivial and nontrivial PCs. To demonstrate topological edge states, a supercell composed of 8 unit cells of NP1 and 8 unit cells of TP is constructed with a domain wall as indicated in Fig. 2b, and is simulated in the COMSOL multiphysics software. Figure 2a presents the projected band diagram. It can be seen that one nearly flat edge state occurs inside the overlapped bandgap of NP1 and TP. The mode profile at $\Gamma$ (Fig. 2b) reveals that the electric field is highly localized at the interface and is decayed exponentially into the bulk. The group index for this edge state is larger than 60, indicating the slow light propagation which has many potential applications in nonlinear optics [34]. However, such a gapped 1D edge state is fundamentally different from the gapless 1D edge state in topological photonic crystals based on the quantum spin hall effects (QSHE) [9, 11].

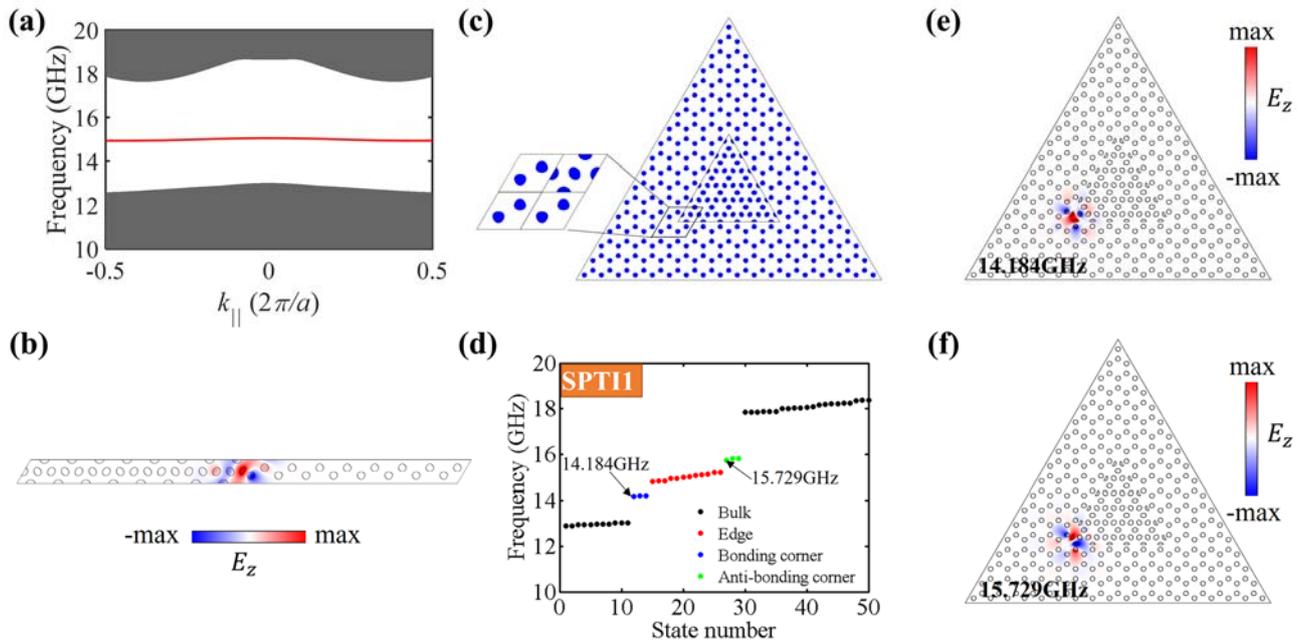

FIG. 2. (a) The bandgap diagram for the supercell with a domain wall between 8 NP1s and 8 TPs as shown in (b); (b) The distribution of the electric field for a 1D edge state at $\Gamma$; (c) The metastructure constructed by NP1s and TPs for the calculation of corner states; (d) Eigenfrequencies of the metastructure in (c); (e) The distribution of the electric field for a bonding corner state; (f) The distribution of the electric field for an anti-bonding corner state.

Figure. 2(e) and 2(f) show the electric field distribution for the bonding corner state and the anti-bonding corner state, respectively. They clearly show that the electric field is highly localized symmetrically or anti-symmetrically at the corner. The gapped edge state is also prerequisite for the realization of a 0D corner state [17, 18]. In order to confirm the existence of 0D corner state, we construct a triangle metastructures as shown in Fig. 2(c), with NP1 at the center and TP at the surround. Figure. 2(d) displays the eigenfrequencies of the metastructures. Different from the existing SPTIs [15-19] with bonding corner states only, we observe three near-degenerate bonding corner states (blue dots) and three anti-bonding corner states (green dots). These in-gap bonding and anti-bonding corner states emerge at the frequencies between the 1D edge states (red dots) and the bulk bands (black dots).

## 4. Tuning frequencies of the edge and corner states

By specifying different frequencies of quadrupolar modes in topology optimization, various optimized nontrivial PCs can be achieved. Figure 3(a) shows four different nontrivial PCs by specifying the frequency of quadrupolar modes at 13.2GHz, 13.5GHz, 14.1GHz and 14.55GHz, respectively. These optimized nontrivial PCs are termed as NP2, NP3, NP4 and NP5, respectively. Thus, five localized states (SPTI1, SPTI2, SPTI3, SPTI4 and SPTI5) appear assembling these optimized nontrivial PCs with TP. Figure. 3(b) shows superimposed projected band diagrams for the supercells of SPTI1, SPTI3, SPTI4 and SPTI5, demonstrating that the edge states of these SPTIs occur at different frequencies. These edge states occur at the overlapped bandgap between 14.526 GHz and 17.625 GHz. Figures. 3(c) and (d) show eigenfrequencies of the metastructures, which are similar to the one shown in Fig. 2c except for replacing NP1 with NP2 and NP3, respectively. As follows from Fig. 2d, Fig 3c and Fig. 3d that the SPTI1, SPTI2 and SPTI3 metastructures have corner states at different frequencies.

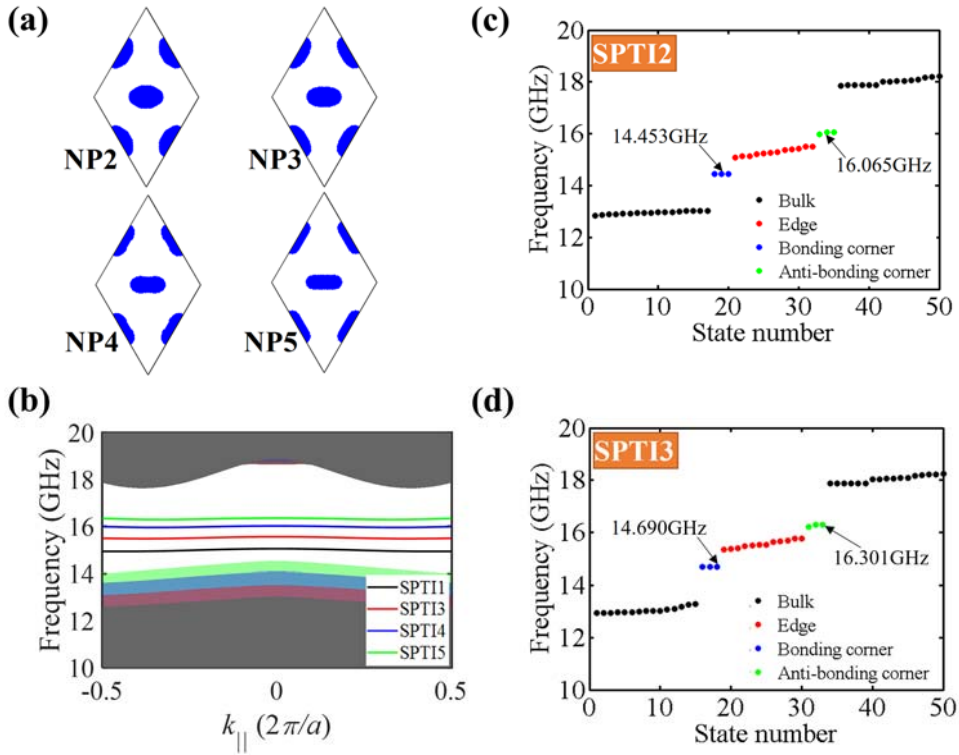

FIG. 3. (a) Optimized nontrivial PCs by specifying the frequency of quadrupolar modes at 13.2GHz, 13.5GHz, 14.1GHz and 14.55GHz, respectively; (b) Superimposed projected band diagrams for SPTI1, SPTI3, SPTI4 and SPTI5; (c) Eigenfrequencies of the metastructures as shown in Fig. 2(c) by replacing NP1 with NP2. (d) Eigenfrequencies of the metastructures as shown in Fig. 2(c) by replacing NP1 with NP3.

*Programming optimized SPTIs for routing topological edge and corner states.* — According to the analysis of topological edge and corner states in the previous section, routing topological edge and corner states with different frequencies can be realized by purposely programming these optimized NPs and TP. As an example, Figure 4a shows the programming diagram of a photonic device composed of NP1, NP3, NP4, NP5 and two pieces of TPs with six channels (boundaries), which aims to operate light waves with frequencies ranging from 14.526 GHz to 17.625 GHz. Since the supercell of two nontrivial PCs exhibits the band gap without any edge states (see the Supplemental Material, Sec. 2), two channels between NP1 and NP5, NP3 and NP4 are blind. Other four channels between nontrivial and trivial PCs, as highlighted with color lines, will be utilized for the propagation of light waves due to the existence of edge states. As shown in Fig. 3b, these four channels support edge states at different frequencies, which do not interfere with each other. As a result, a point source with different frequencies at the center of the device, as denoted by the red asterisk in Fig. 4a, may propagate along different channels. Figures 4b-e show the distributions of the electric field when the source having the frequency 15.06 GHz, 15.51 GHz, 16.09 GHz and 16.37 GHz, respectively. It can be clearly seen that topological edge states with different frequencies are routed into different channels without crosstalk.

Figure 5(a) illustrates the schematic diagram of another programming diagram of a photonic device for routing topological corner states. The device consists of three different nontrivial PCs, NP1, NP2 and NP3, surrounded by TP. Three 60-deg corners are formed between each nontrivial PC and trivial PC, being named as C1, C2 and C3, respectively. These corners possess different frequency of the bonding corner state, as shown in Fig. 2 (d), Fig. 3(c) and Fig. 3(d). We put a point source at the center of the internal triangle region for nontrivial PCs, as denoted by the red asterisk. Figures 5(b-d) show the simulated electric field distribution for the point source with the frequency being 14.184 GHz, 14.453 GHz and 14.690 GHz, corresponding to the frequency of the bonding corner state for each corner, respectively.

Thus, the localized state of each corner can be excited selectively by changing the frequency of the point source. Therefore, the corner states with different frequencies are routed into different corners without crosstalk. Routing anti-bonding topological corner states is demonstrated in Supplemental Material Sec. 3. Besides, we demonstrate the robustness of corner states against perturbations in Supplemental Material Sec. 4, confirming that the topological corner state possesses a good immunity against defect.

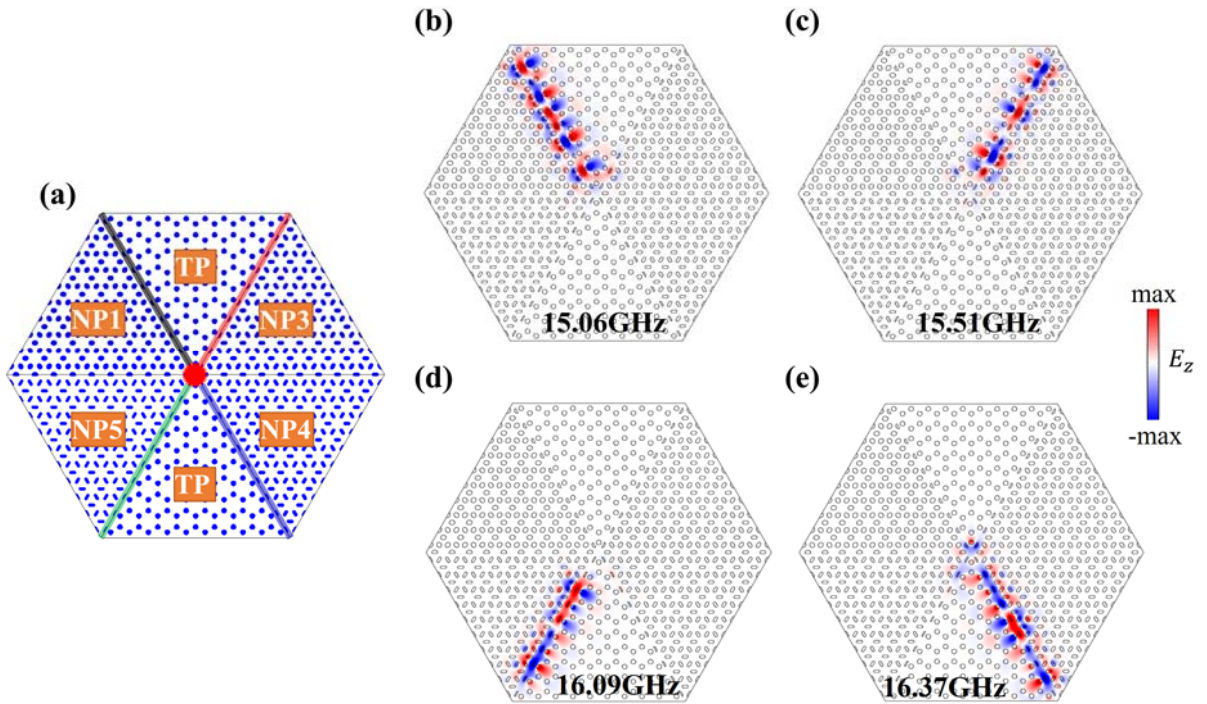

FIG. 4. Routing topological edge states with different frequencies. (a) Schematic of the photonic device for routing topological edge states. The color lines denote four channels formed between non-trivial PCs and trivial PCs. (b), (c), (d), (e) Simulated electric field distribution for the point source with the frequency being 15.06 GHz, 15.51 GHz, 16.09 GHz and 16.37 GHz, respectively.

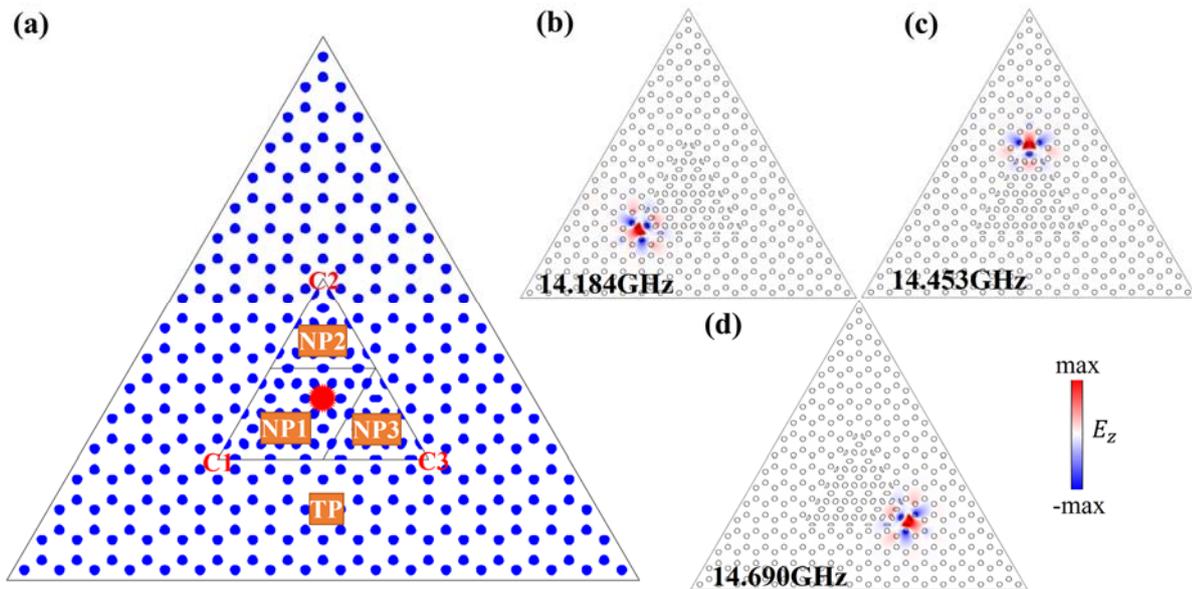

FIG. 5. Routing topological corner states with different frequencies. (a) Schematic of the photonic device for routing topological corner states. (b), (c), (d) Simulated electric field distribution for the point source with the frequency being 14.184 GHz, 14.453 GHz and 14.690 GHz, respectively.

## 5. Conclusions

We have proposed an intelligent numerical method for inverse design of higher-order photonic topological insulators. For the second-order topological insulators, the optimized structures support a nearly flat edge state and highly localized corner states, which may have the potential application in enhancing light-matter interaction. Rather than traditional states, which only support bonding corner states, the optimized topological states support both bonding and anti-bonding corner states. In particular, by inversely designing several types of topological structures with different frequencies for both edge and corner states and purposely programming their properties, we have realized routing topological edge and corner states with different frequencies. Our findings provide great flexibilities for utilizing topological edge and corner states, which is important for developing photonic devices with novel functionalities.

**Acknowledgements**

The work was supported by the Australian Research Council (grants FT130101094 and DP200101168). Yafeng Chen acknowledges a support from China Scholarship Council.